\documentclass[sigconf]{acmart}
\usepackage{textgreek}
\usepackage{algorithm}
\usepackage{algpseudocode}
\usepackage{subcaption}
\AtBeginDocument{%
  }

\begin{document}

\title{Categorising SME Bank Transactions with Machine Learning and Synthetic Data Generation}


\author{Pietro Alessandro Aluffi}
\authornote{Corresponding Author}
\email{pietro.aluffi@warwick.ac.uk}
\affiliation{%
  \institution{University of Warwick}
  \country{}
}
\affiliation{%
  \institution{Navrisk}
  \country{}
}

\author{Brandi Jess}
\affiliation{%
  \institution{Navrisk}
  \country{}
}

\author{Marya Bazzi}
\affiliation{%
  \institution{University of Warwick}
  \country{}
}
\affiliation{%
  \institution{SME Capital}
  \country{}
}
\affiliation{%
  \institution{sea.dev}
  \country{}
}

\author{Kate Kennedy}
\affiliation{%
  \institution{SME Capital}
  \country{}
}
\affiliation{%
  \institution{Navrisk}
  \country{}
}

\author{Matt Arderne}
\affiliation{%
  \institution{SME Capital}
  \country{}
}
\affiliation{%
  \institution{sea.dev}
  \country{}
}

\author{Daniel Rodrigues}
\affiliation{%
  \institution{SME Capital}
  \country{}
}
\affiliation{%
  \institution{Navrisk}
  \country{}
}

\author{Martin Lotz}
\affiliation{%
  \institution{University of Warwick}
  \country{}
}

\begin{abstract}
Despite their significant economic contributions, Small and Medium Enterprises (SMEs) face persistent barriers to securing traditional financing due to information asymmetries. Cash flow lending has emerged as a promising alternative, but its effectiveness depends on accurate modelling of transaction-level data. The main challenge in SME transaction analysis lies in the unstructured nature of textual descriptions, characterised by extreme abbreviations, limited context, and imbalanced label distributions. While consumer transaction descriptions often show significant commonalities across individuals, SME transaction descriptions are typically nonstandard and inconsistent across businesses and industries. To address some of these challenges, we propose a bank categorisation pipeline that leverages synthetic data generation to augment existing transaction data sets. Our approach comprises three core components: (1) a synthetic data generation module that replicates transaction properties while preserving context and semantic meaning; (2) a fine-tuned classification model trained on this enriched dataset; and (3) a calibration methodology that aligns model outputs with real-world label distributions. Experimental results demonstrate that our approach achieves 73.49\% (±5.09) standard accuracy on held-out data, with high-confidence predictions reaching 90.36\% (±6.52) accuracy. The model exhibits robust generalisation across different types of SMEs and transactions, which makes it suitable for practical deployment in cash-flow lending applications. By addressing core data challenges, namely, scarcity, noise, and imbalance, our framework provides a practical solution to build robust classification systems in data-sparse SME lending contexts.
\end{abstract}

%

\keywords{SME transaction classification, synthetic data generation, financial text classification, calibrated classification, Open Banking, cash flow lending}


\maketitle

\section{Introduction}

The digital transformation of financial services has created new opportunities for data-driven access to finance and credit assessment, particularly for sectors that do not have easy access to traditional financial instruments, such as Small and Medium-sized Enterprises (SMEs). 
Despite their substantial contributions to innovation, employment, and GDP, SMEs continue to face major barriers in securing traditional bank financing~\citep{gherghina_small_2020, beck_small_2006}. Banks often perceive SMEs as high-risk entities due to incomplete, inaccurate, out-of-date, or non-standardised financial data.
Due to their limited financial buffers and narrower market focus, SMEs often fail quickly and traditional credit assessments. 
become insufficient for capturing and reacting to financial distress in a timely fashion.

Leveraging future cash flows, cash flow lending has emerged as a promising alternative to assess creditworthiness for assets-light businesses~\cite{ndifor_effect_2023}. The effectiveness of cash flow lending depends on accurate modelling and interpretation of transaction-level cash flow data, both at the underwriting stage and throughout the loan lifecycle. Despite the advent of the Open Banking framework that improves data availability, granularity, and transparency, challenges related to data processing and interpretation remain: while consumer financial activities tend to be more homogeneous from one individual to another and use established methods for classification, SME transactions are often sparse, unclear, and highly dependent on their specific context \cite{hossain_large-scale_2020}. SME transaction data can vary both in structure and semantics between different businesses and sectors, which complicates the extraction of consistent insights from raw data. 



The core challenge in automatically classifying SME bank transactions lies in the scarcity and unstructured nature of their textual descriptions, characterised by extreme short-hand abbreviations, limited contextual information, and highly imbalanced label distributions~\cite{ kotios_deep_2022,ta_specialized_2023}. These challenges make conventional classification methods difficult to generalise. Although manual annotations or rule-based heuristics provide some interpretability and domain alignment~\cite{kotios_deep_2022}, these approaches struggle to scale and adapt to the evolving and heterogeneous nature of SME financial data. In practice, resource constraints on manual labelling and inference further limit the feasibility of large and complex classification systems. Therefore, new approaches must be able to generalise from small samples and better recognise different transaction types for a given SME.

We propose a bank transaction categorisation pipeline\footnote{Pipeline implementation code available upon request.} that leverages synthetic data generation to augment existing SME transaction datasets. 
Synthetic data provides a scalable and privacy-preserving approach to simulate realistic transaction patterns, even in the presence of limited or highly varied data where meaningful behaviours are often under-represented. Our pipeline comprises three core components: (1) a synthetic data generation module that replicates SME transaction properties while preserving contextual realism; (2) a fine-tuned and calibrated categorisation model trained on this enriched dataset; and (3) an evaluation on manually labelled transactions.

\section{Related Work}
Following the financial crisis, UK banks reduced SME lending, creating a £95 billion finance gap (2015-2022) filled by challenger banks and alternative finance providers \cite{house_of_commons_treasury_committee_sme_2024}.
Cash flow lending requires robust risk modelling from transaction histories. Misclassification can cause adverse selection or default, making accurate categorization critical. 
SME transaction analyses hereby face major challenges. Bank descriptions are short, noisy, and inconsistent~\cite{toran_scalable_2023}. Weak supervision combining rule-based labelling with neural networks, and CNN/RNN models for pattern detection attempts to address these unstructured descriptions~\cite{banu_financial_2024}. Inconsistent naming, including abbreviations, also reduces NLP effectiveness~\cite{garcia-mendez_demographic_2021, schoinas_normalisation_2019}. Metadata integration~\cite{toran_scalable_2023, garcia-mendez_identifying_2024}, specialised tokenization~\cite{ta_specialized_2023}, and hybrid architectures are common approaches to address these challenges~\cite{hjelkrem_explaining_2023}. In addition, manual annotation is costly for domain-specific SME categorization~\cite{zhang_credit_2020}, which can be partially resolved with fine-tuned BERT and zero-shot classification for unlabelled data~\cite{masoumi_natural_2024}. These limited state-of-the-art approaches for categorising SME bank transactions, including LSTMs with anomaly detection \cite{hussain_ai-enhanced_2023}, end-to-end learning systems\cite{masud_practical_2008,shao_anomaly_2021}, and LLM-enabled synthetic transaction generation and zero-shot classification~\cite{he_annollm_2024, li_large_2023} facilitate some dynamic training for SME variability; circular dependency persists: classification needs context, context models need labels~\cite{kotios_deep_2022}. The following main limitations remain: one, the reliance on extensive manual annotations~\cite{zhang_credit_2020}. Two, limited generalisability to unseen transactions (e.g., a given SME is highly volatile with multiple temporal drifts in transaction patterns throughout its lifetime) or SMEs~\cite{toran_scalable_2023}.


Our key contribution is two-fold: First, we develop a generalisable, robust pipeline from a sparse set of initial manual annotations. Second, our pipeline leverages LLMs for synthetic data generation in order to a) mitigate data scarcity
and, more importantly, b) cater to SME context dependency by amplifying business-specific  idiosyncrasies as part of the categorisation pipeline. Our contribution thus improves performance on under-represented or emerging transaction types. 

The setup of our pipeline is as follows: (1) generating class-balanced data via LLM prompting, (2) fine-tuning transformers with focal loss, and (3) calibrating outputs against real-world distributions.



\section{Methodology}
\subsection{Problem Formulation}

We define our classification task based on financial transactions. Each data point comprises a transaction described by free-text fields and associated metadata. The structure of our dataset is formally defined as follows.

\begin{definition}[Dataset]
\label{def:dataset}
The dataset is defined as \( T(D, L) \), where:
\begin{itemize}
  \item \( D = \{d_1, d_2, \dots, d_n\} \) is the set of cleaned transaction descriptions.
  \item \( L = \{l_1, l_2, \dots, l_n\} \) is the set of corresponding manually assigned labels from a finite set of predefined transaction categories indicating the transaction type.
\end{itemize}
\end{definition}

Before the classification task, we apply specific preprocessing steps to standardise descriptions and aggregate similar entries. These steps are encapsulated in the following functions:

\begin{definition}[Preprocessing Functions]
\label{def:preprocessing}
The preprocessing involves:
\begin{itemize}
  \item \(\textsc{Clean}(\cdot)\): Standardises and cleans raw transaction descriptions \(d_{raw}\) to produce \(d \in D\).
  \item \(\textsc{Group}(\cdot)\): Aggregates semantically similar transaction entries within \(D\).
\end{itemize}
\end{definition}

Given the potential scarcity of labelled data, we augment the training set using synthetic examples. This process is defined as:

\begin{definition}[Data Augmentation]
\label{def:augmentation}
To address limited labelled data, we apply \(\textsc{Generate}(\cdot)\), a function that synthesises new labelled examples (\(d', l'\)) based on the existing \(T(D, L)\) to augment the training set.
\end{definition}

Finally, classification performed by a machine learning model, specifically a fine-tuned language model chosen for its suitability to financial text:

\begin{definition}[Classification Model and Calibration]
\label{def:classifier}
The classification process involves two main stages:
\begin{itemize}
    \item \textbf{Fine-tuning:} The core classification function \(\textsc{Finetune}(\cdot)\) is obtained by fine-tuning \textsc{FinBERT} \cite{araci_finbert_2019}, a domain-specific language model pre-trained on financial texts. This function \(\textsc{Finetune}\) maps a preprocessed transaction description \(d_i\) to raw output logits over the set of possible categories \(L_{cat}\).
    \item \textbf{Calibration:} To ensure the model's output probabilities are well-calibrated and align with observed data distributions, a subsequent calibration function, denoted \(\textsc{Calibrate}(\cdot)\), is applied to the logits produced by \(\textsc{Finetune}(\cdot)\). This function implements temperature scaling \cite{guo_calibration_2017}.
\end{itemize}
The application of \(\textsc{Calibrate}(\cdot)\) to the output of \(\textsc{Finetune}(\cdot)\) yields the final calibrated classifier, which produces the probability distribution \(P(L|d_i)\).
\end{definition}

Data collection, pre-processing to obtain \(D\), and labelling processes for \(L\) are described in Sections~\ref{labelling} and~\ref{preprocessing}.
Implementation details for the data augmentation methodology can be found in Section~\ref{synthetic}.
Implementation details for classification and calibration are in Section~\ref{fine_tuning} and \ref{calibration}.

\begin{algorithm}[H] 
\caption{Transaction Classification Pipeline. This process utilises functions for cleaning (\textsc{Clean}) and grouping (\textsc{Group}), defined in Def.~\ref{def:preprocessing}, data augmentation (\textsc{Generate}), defined in Def.~\ref{def:augmentation}, and the classification model function (\textsc{Finetune}) followed by calibration (\textsc{Calibrate}), both described in Def.~\ref{def:classifier}.}
\label{alg:pipeline}
\begin{algorithmic}[1]
\Require Dataset $T(D, L)$ with transactions $D = \{d_1, \dots, d_n\}$ and labels $L = \{l_1, \dots, l_n\}$
\Ensure Trained and calibrated classifier function $f'(\cdot)$ representing the final model output.

\State $D' \gets \textsc{Clean}(D)$ 
\State $D'' \gets \textsc{Group}(D')$ 
\State $T_{\text{aug}}(D_{\text{aug}}, L_{\text{aug}}) \gets \textsc{Generate}(T(D'', L))$
\State $f(\cdot) \gets \textsc{Finetune}(T_{\text{aug}})$ 
\State $f'(\cdot) \gets \textsc{Calibrate}(f(\cdot))$ 
\State \textbf{Evaluation:}
\Statex \hspace{\algorithmicindent} Use data from SMEs 1 \& 2 for training and validation sets.
\Statex \hspace{\algorithmicindent} Use data from SME 3 for out-of-sample test evaluation.
\Statex \hspace{\algorithmicindent} Perform manual expert review on model predictions for unlabelled data originating from six distinct SMEs.
\end{algorithmic}
\end{algorithm}

The dataset was obtained through the Open Banking protocol from our industry partner that provides loans to SMEs throughout the UK, comprising transaction records from nine SMEs in the manufacturing sector (as defined by Companies House condensed SIC codes)\footnote{\url{https://resources.companieshouse.gov.uk/sic/}} selected to represent various business models and banking providers. An illustrative example of the dataset is provided in Table~\ref{tab:transaction_example}. For training and evaluation, we use transaction data from three of these nine SMEs, for which manually annotated ground truth labels are available. Data from two of these firms are used for model training and validation, using these ground truth labels. The third firm, also with ground truth labels, is held out entirely for out-of-sample evaluation, where its labels are used alongside standard classification metrics. To further assess the generalisation of the model and perform additional validation, we use data from the remaining six SMEs. Table~\ref{tab:transaction_data_new_style} summarises the temporal coverage and transaction volume of these firms. For these firms, domain experts from our industry partner conduct validation checks by manually annotating 100 transactions chosen uniformly at random across the timeframe for each, allowing model predictions to be compared against these expert annotations.

\begin{table}[h]
    \centering
    \caption{\textbf{Example of Open Banking Transaction Data}}
    \begin{tabular}{l|r|l}
        \toprule
        \textbf{Date} & \textbf{Amount (£)} & \textbf{Description} \\
        \hline
        2024-03-01 & 5,200.00 & ABC SUPPLIERS LTD INV12345 DD \\
        2024-03-05 & 850.75 & UTILTY ENERG PAY MAR2024 9876 FT\\
        2024-03-10 & 12,000.00 & PAYROLL 0456 BULKPAY\\
        2024-03-15 & 2,300.50 & XYZ TRANSPORT INC 2024-987 BACS \\
        \bottomrule
    \end{tabular}
    \label{tab:transaction_example}
\end{table}

\begin{table}[h] 
    \centering
    \caption{\textbf{Transaction Data Summary}} 
    \label{tab:transaction_data_new_style}
    \begin{tabular}{l|c|c|r|r} 
        \toprule
        \textbf{Name} & \textbf{Start Date} & \textbf{End Date} & \textbf{Transactions} & \textbf{Total Days} \\
        \hline 
        company1 & 2022-07-26 & 2024-07-29 & 2984 & 734 \\
        company2 & 2023-10-10 & 2024-09-30 & 3660 & 356 \\
        company3 & 2022-07-04 & 2024-09-30 & 8238 & 819 \\
        company4 & 2022-07-13 & 2024-09-30 & 5689 & 810 \\
        company5 & 2022-07-19 & 2024-09-30 & 2726 & 804 \\
        company6 & 2022-06-27 & 2024-09-27 & 9884 & 823 \\
        \bottomrule
    \end{tabular}
\end{table}
\subsubsection{Manual Labelling} \label{labelling}

Accurately labelled transaction data form the foundation of our classification model. The process began with domain experts from our industry partner, who have a deep understanding of business models and operational intricacies of the companies included in the dataset. They assigned each transaction a label from a predefined set of categories, as detailed in Table~\ref{tab:label_dist_comparison}. These manual annotations provided the ground truth labels for training. In addition to labelling data for training, the domain experts also labelled data to validate the classification model. That is, for a subset of data from additional SMEs, a random sample of transactions was manually annotated to assess model generalisation. While this labelling process is time intensive, having domain experts label data is vital for establishing a reliable benchmark for our automated categorisation pipeline, given their deep engagement with and understanding of the SMEs.

\subsubsection{Preprocessing} \label{preprocessing}

The proposed preprocessing pipeline transforms raw transaction descriptions into standardised textual representations suitable for semantic labelling and categorisation. Let $D = \{d_1, d_2, \dots, d_n\}$ denote the set of raw transaction descriptions. Each $d_i \in D$ is first passed through a cleaning function $\textsc{Clean}(\cdot)$, which applies a series of text normalization steps: (1) replacement of common financial abbreviations (e.g., ``ATM'' $\rightarrow$ ``cash'', ``BACS'' $\rightarrow$ ``debit''), (2) conversion to lowercase, (3) removal of punctuation and irrelevant characters using regular expressions, (4) filtering of purely numeric or non-informative tokens (e.g., reference numbers), and (5) removal of stop words and domain-specific terms (e.g., ``ref'', ``ltd'', month abbreviations). If the cleaned result is empty, it is replaced with a placeholder token (e.g., ``nodescription''), which is discarded in downstream steps.

After cleaning, we apply a grouping function $\textsc{Group}(\cdot)$ that groups semantically equivalent cleaned descriptions. This allows variations of a transaction, such as \texttt{``PYMT inv 24534 AMZN''} and \texttt{``PYMT inv 234325 AMAZON''}, to be reduced to a single form (e.g., \texttt{``amazon payment''}). Once a label is assigned to the cleaned form, it can be assigned to all transactions in the group, facilitating consistent labelling across similar transactions. This approach enables scalable manual annotation and robust downstream classification. Our preprocessing procedure builds on the pipeline introduced in~\citet{toran_scalable_2023}, adapted here for purely text-based analysis.

\subsubsection{Synthetic Data Augmentation}\label{synthetic}

We define the synthetic transaction generation function $\textsc{Generate}(d, c, n)$, where $d$ is a transaction description, $c$ is its associated category, and $n$ is the number of synthetic samples to generate. This function is used to augment the labelled dataset with realistic and semantically consistent variations of $d$, especially for under-represented categories.

Synthetic descriptions are generated using \texttt{gpt-4o} via the OpenAI API, with temperature set to 0.7 and a maximum of 512 tokens per request. The prompts are designed to rephrase the original transaction while maintaining semantic meaning and contextual relevance to the associated category. For instance, an original transaction description like:
\begin{quote}
\texttt{biffa waste servic ltd b47391 bbp}
\end{quote}
could yield synthetic variations such as:
\begin{itemize}
    \item \texttt{‘veolia refuse service payment ref ltd vrs b47392’}
    \item \texttt{‘suez disposal services ltd payment ref sd b47393’}
    \item \texttt{‘grundon rubbish collection fee ref ltd grc b47395’}
\end{itemize}

The number of synthetic samples $n$ per class is determined using inverse frequency scaling, increasing the representation of minority classes to approximate a more balanced, though not perfectly uniform, class distribution. All generated samples are post-processed using the same $\textsc{Clean}(\cdot)$ function described in Section~\ref{preprocessing}. Domain experts performed manual validation to verify that the generated outputs were realistic, coherent, and category-consistent. However, we emphasise that this augmentation step was exploratory: we did not over-optimise prompt engineering, filtering, or model parameters. The objective of this study was to assess whether the generation of basic, semantically guided synthetic data could improve classification performance, not to build an optimised generation pipeline. More complex augmentation strategies remain a direction for future work.

\subsubsection{Fine-tuning}\label{fine_tuning}

We define the function \textsc{FineTuneFinBERT}$(S)$, where $S = \{(d_i, l_i)\}_{i=1}^n$, as the balanced synthetic data set consisting only of augmented and preprocessed transaction descriptions and their corresponding labels. The objective is to learn a classifier $f(\cdot)$ by fine-tuning a domain-specific language model on $S$, where each $d_i$ is a preprocessed input and each $l_i$ is drawn from the label set $L$. The model $f$ is initialised as a pre-trained FinBERT~\cite{araci_finbert_2019} encoder with a classification head adapted for $|L|$ output classes. Fine-tuning is performed using a weighted focal loss function~\cite{lin_focal_2018}, which combines class weighting with the focal mechanism to mitigate the effects of class imbalance and over-confident predictions. Let $p_t$ denote the predicted probability of the model for the true class $t$, and let $\alpha_t$ be the weight associated with class $t$. The weighted focal loss $\mathcal{L}_{\text{focal}}$ for a single instance is given by:

\[
\mathcal{L}_{\text{focal}} = -\alpha_t (1 - p_t)^\gamma \log(p_t)
\]
where $\gamma \geq 0$ is a focusing parameter (commonly $\gamma = 2$) that downweights the loss assigned to well-classified examples, placing more emphasis on hard or misclassified instances. We implement focal loss over standard cross-entropy loss or unweighted alternatives because of the label distribution and semantic similarity in free-text descriptions. In this case, the model may become overconfident in its predictions even when incorrect. Focal loss directly addresses these issues by reducing the contribution of easy, high-confidence examples and amplifying the importance of harder cases, thus promoting a more balanced classifier. Class weights $\alpha_t$ are computed using inverse frequency statistics from the training set, ensuring that under-represented classes receive proportionally greater emphasis during training. The dataset $S$ is stratified into training and validation subsets, tokenised using the FinBERT tokenizer, and encoded with truncation and padding. 

Table~\ref{tab:hyperparameters} summarises the hyperparameters and training configuration used across all experiments, including the baseline methods for comparison.

\begin{table}[h!]
\centering
\caption{Model Hyperparameters and Training Configuration}
\label{tab:hyperparameters}
\begin{tabular}{@{}llc@{}}
\toprule
\textbf{Component} & \textbf{Parameter} & \textbf{Value} \\
\midrule
{\textbf{FinBERT Fine-tuning}}& Base Model & ProsusAI/finbert \\
& Learning Rate & 2e-5 \\
& Batch Size & 16 \\
& Max Sequence Length & 256 \\
& Epochs & 3 \\
& Warmup Steps & 500 \\
\midrule
{\textbf{TF-IDF Models}}& Max Features & 10,000 \\
& N-gram Range & (1, 2) \\
& Stop Words & English \\
& Class Weight & Balanced \\
\midrule
{\textbf{Random Forest}}& N Estimators & 100 \\
& Random State & 42 \\
& Class Weight & Balanced \\
\midrule
{\textbf{Logistic Regression}}& Max Iterations & 1,000 \\
& Random State & 42 \\
& Class Weight & Balanced \\
\midrule
\bottomrule
\end{tabular}
\end{table}

The output of FINETUNE($\cdot$) is the fine-tuned model $\hat{f}$, stored for downstream calibration and inference. 

\subsubsection{Calibration}\label{calibration}

While the fine-tuning stage aims to improve classification performance, especially for under-represented classes, oversampling and augmentation can hide from the model the original class distribution. However, in practice, the real-world frequency of transaction labels is inherently specific to the type of businesses and the sector they operate in. For example, a business specialising in power engineering (suppliers) may have many payments to contractors, which are semantically similar to energy transactions (utilities). Without correction, these transactions may be misclassified as utilities, a category that typically occurs with low frequency in most businesses. To avoid this, we implement a calibration step to adjust predicted probabilities so they more accurately reflect the true distribution of transaction types observed in operational settings. We define the calibration procedure \textsc{CALIBRATE}$(\hat{f}, D_\text{real})$, where $\hat{f}$ is the fine-tuned classifier from the previous step and $D_\text{real} = \{(d_i, l_i)\}_{i=1}^m$ is a labelled dataset consisting of real transaction data from the two manually labelled companies and their ground truth labels. The objective is to align the predictive confidence and label distribution of $\hat{f}$ with those observed in real-world data.

Given predicted logits from $\hat{f}$, we apply temperature scaling~\cite{guo_calibration_2017} to calibrate the output probabilities. Let $\mathbf{z}_i \in \mathbb{R}^{|L|}$ be the pre-softmax logits, for instance $i$, and let $T \in \mathbb{R}^{+}$ be a learnt temperature parameter. The calibrated logits $\tilde{\mathbf{z}}_i$ are computed as:

\[
\tilde{\mathbf{z}}_i = \frac{\mathbf{z}_i}{T} + \mathbf{b}
\]

where $\mathbf{b} \in \mathbb{R}^{|L|}$ is a learned bias term. The parameters $(T, \mathbf{b})$ are optimised to minimise the negative log-likelihood (NLL) on a held-out calibration set. The result is a calibrated model whose output probabilities better reflect both predictive confidence and the expected distribution of transaction categories in deployment. Calibration effectiveness is evaluated using metrics such as expected calibration error (ECE) and NLL.

\subsubsection{Evaluation}

We evaluated the calibrated classifier $\hat{f}$ using a 5-fold cross-validation on a labelled dataset of real transaction descriptions and their corresponding ground truth labels.
For each transaction $d_i$, the model outputs a calibrated class probability distribution via:

\[
\hat{P}'(y|d_i) = \textsc{CALIBRATE}(\hat{f}(d_i))
\]

where $\textsc{CALIBRATE}(\cdot)$ denotes temperature scaling, applied to the uncalibrated logits produced by the base classifier $\hat{f}$. The final predicted label and associated confidence score for each input are computed as:
\[
\hat{l}_i = \arg\max_y \hat{P}'(y|d_i)
\quad \text{and} \quad
\text{conf}_i = \max_y \hat{P}'(y|d_i)
\]

We report multiple evaluation metrics:
\begin{itemize}
    \item \textbf{Standard Accuracy}: Proportion of correct predictions across all test instances.
    \item \textbf{High-Confidence Accuracy}: Accuracy computed on the samples where $\text{conf}_i > 0.8$.
    \item \textbf{Top-Class Confidence Accuracy}: Accuracy among the top 10\% most confident predictions per fold.
    \item \textbf{Top-2 Accuracy}: Fraction of instances where the true label appears within the two predicted top classes: \[
    \hat{l}_i \in \text{Top-2}(\hat{P}'(y|d_i))
    \]
\end{itemize}

Following cross-validation, the model was retrained on the full labelled dataset and applied to unlabelled transaction data to generate probabilistic label predictions. These outputs can be used for downstream tasks such as weak supervision, anomaly detection, or prioritised human review.

\section{Results}

We evaluated both individual components and the end-to-end performance of the suggested pipeline. We begin by validating our synthetic data generation approach, followed by calibration, classification performance on held-out data, and comparative analysis against baseline methods.
\subsection{Synthetic Data Quality Evaluation}

\paragraph{Statistical and Linguistic Properties}
Our synthetic data closely matches the linguistic characteristics of real transactions. Length distributions show similar means (real: 36.3±19.7 vs. synthetic: 36.8±17.7 characters). While the synthetic vocabulary expanded significantly (6,576 vs. 1,416 tokens), it maintains 48.0\% coverage of the original vocabulary with a Jaccard similarity of 0.093. This expansion is desirable as it introduces linguistic variation while preserving domain-relevant terms. 

\paragraph{Semantic Coherence and Diversity}
Semantic analysis using BERT \cite{devlin_bert_2019} embeddings reveals a strong alignment between real and synthetic data. The mean cosine similarity of 0.879 (±0.048) demonstrates that synthetic transactions preserve semantic meaning within their assigned categories. Importantly, our generation process maintains diversity with 94.2\% unique synthetic examples and a diversity score of 0.167, avoiding mode collapse. Category coherence scores remain consistently high across all classes (0.835-0.897).

\paragraph{Class Balancing Strategy}
Our synthetic augmentation employs an inverse-frequency scaling strategy to address class imbalance:

\begin{table}[h]
\centering
\small
\begin{tabular}{lrrc}
\hline
\textbf{Category} & \textbf{Real} & \textbf{Synthetic} & \textbf{Ratio} \\
\hline
Suppliers & 565 & 565 & 1.0× \\
Payroll/Consultants & 460 & 460 & 1.0× \\
Sundries & 177 & 1,062 & 6.0× \\
Software/IT & 160 & 960 & 6.0× \\
Travel & 137 & 959 & 7.0× \\
Tax & 104 & 1,040 & 10.0× \\
Utilities & 97 & 988 & 10.2× \\
Marketing & 84 & 840 & 10.0× \\
Inventory & 52 & 936 & 18.0× \\
Debt/Loan & 34 & 952 & 28.0× \\
Rent & 27 & 810 & 30.0× \\
\hline
\end{tabular}
\caption{Synthetic data generation ratios by category, showing inverse-frequency scaling to address class imbalance.}
\label{tab:synthetic_ratios}
\end{table}

This strategy generates up to 30× synthetic examples for minority classes while maintaining 1:1 ratios for majority classes, effectively balancing the training distribution without overwhelming the model with synthetic data.

\subsection{Calibration Performance}

\begin{figure}[b]
    \captionsetup{skip=10pt} 
    \centering
    \begin{subfigure}[b]{0.7\linewidth}
        \centering
        \includegraphics[width=\linewidth, height=0.3\textheight]{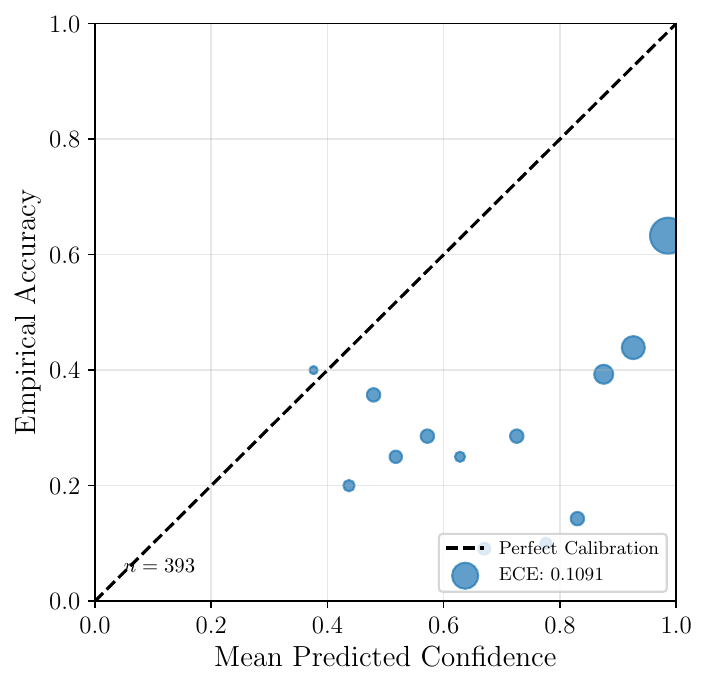}
        \caption{Calibration plot \textbf{before} calibration.}
        \label{fig:calibration1}
    \end{subfigure}

    \vspace{0.5em}

    \begin{subfigure}[b]{0.7\linewidth}
        \centering
        \includegraphics[width=\linewidth, height=0.3\textheight]{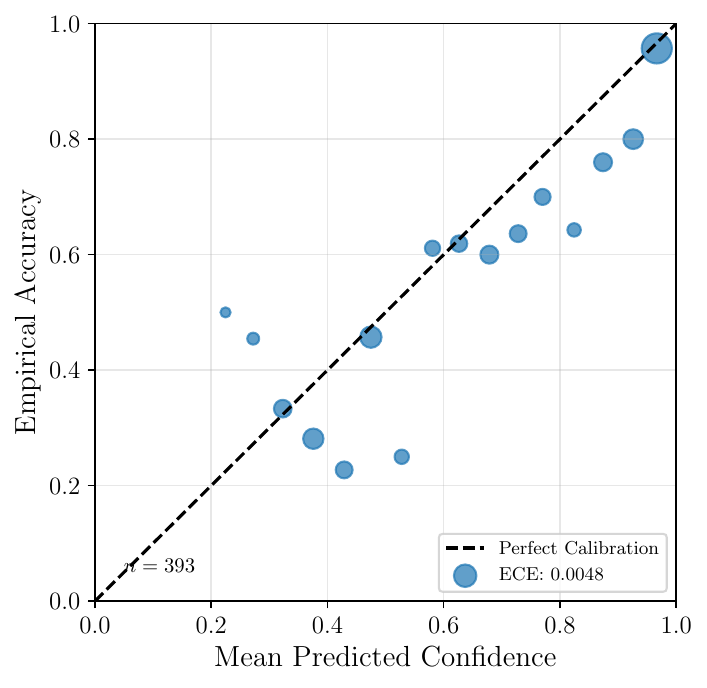}
        \caption{Calibration plot \textbf{after} calibration.}
        \label{fig:calibration2}
    \end{subfigure}

    \caption{Calibration plots showing mean predicted confidence (x-axis) versus empirical accuracy (y-axis) on the test set, before (a) and after (b) applying the calibration method. Points represent binned predictions, with the diagonal line indicating perfect calibration.}
    \label{fig:calibration}
\end{figure}

To evaluate the reliability of the model’s probability estimates, we evaluated the calibrated classifier using standard calibration metrics on a holdout test set. As shown in Figure~\ref{fig:calibration}, the model initially exhibited significant miscalibration, with predicted confidences notably higher than empirical accuracies (top panel). This is quantified by a relatively high ECE of 0.1091 before calibration. After calibration, the alignment between predicted confidence and actual accuracy improved substantially (bottom panel), reducing the ECE to 0.0048, indicating well-calibrated predictions. Furthermore, the NLL was measured at 0.8141, supporting the conclusion that the model produces reasonably calibrated probabilistic outputs. Table~\ref{tab:label_dist_comparison} also shows that calibration improved the alignment between the predicted and true label distributions, which is especially important in class-imbalanced domains such as SME. 

\begin{table}[h]
    \centering
    \caption{\textbf{Label Distribution Comparison}}
    \begin{tabular}{l|c|c}
        \toprule
        \textbf{Label Name} & \textbf{Target } & \textbf{Predicted } \\
        \hline
        charges / fees & 0.0549 & 0.0462 \\
        debt / loan repayment & 0.0122 & 0.0154 \\
        marketing / advertising & 0.0427 & 0.0769 \\
        payroll / consultants & 0.1280 & 0.1077 \\
        rent & 0.0091 & 0.0154 \\
        software / it & 0.0732 & 0.0308 \\
        sundries & 0.0976 & 0.1538 \\
        suppliers & 0.4055 & 0.3231 \\
        tax & 0.0366 & 0.0615 \\
        travel & 0.1159 & 0.1385 \\
        utilities & 0.0244 & 0.0308 \\
        \bottomrule
    \end{tabular}
    \label{tab:label_dist_comparison}
\end{table}

\subsection{Classification Performance}

\begin{figure}[b!]
    \centering
    \includegraphics[width=0.7\linewidth]{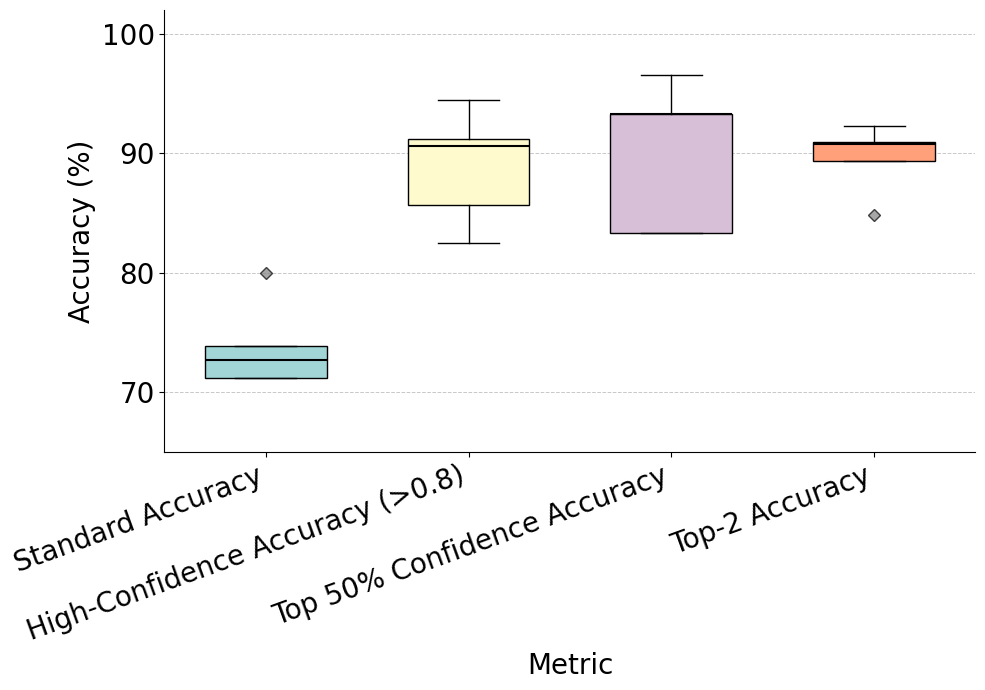}
    \caption{Accuracy distribution across folds for standard, high-confidence, top-50\%, and top-2 predictions.}
    \label{fig:conf_accuracy_boxplot}
\end{figure}

We evaluated the final classification performance using 5-fold cross-validation on the held-out SME. The average standard accuracy is 73.49\% ($\pm$5.09\%), indicating consistent generalisation performance across the divides despite label imbalance and class sparsity. To further assess performance across varying levels of model confidence, we report several additional reliability-aware metrics. Figure~\ref{fig:conf_accuracy_boxplot} illustrates the accuracy distribution across the cross-validation folds for the standard accuracy as well as for these confidence-aware measures. The specific confidence-aware metrics are:

\begin{itemize}
    \item High-Confidence Accuracy (conf \> 0.8\textbf{)}: 90.36\% ($\pm$6.52\%), demonstrating strong precision when the model is confident.
    \item Top 50\% Confidence Accuracy: 88.55\% ($\pm$5.21\%), reflecting model robustness in the most confident half of the predictions, useful for confidence-based filtering.
    \item Top-2 Accuracy: 89.63\% ($\pm$4.51\%), indicating that the correct class appears within the top two predictions in the vast majority of cases, supporting semi-automated review pipelines.
\end{itemize}

These results suggest that the model performs reliably across both low- and high-confidence scenarios, and that confidence estimates can be effectively leveraged to support human-in-the-loop workflows or probabilistic label refinement strategies in noisy financial text classification settings.

\vspace{1em}

\subsection{Comparative Analysis}
To address the need for comprehensive baseline comparisons and evaluate the contribution of individual components in our pipeline, we conducted extensive experiments using 5-fold stratified cross-validation.

\subsubsection{Baseline Methods}
We compared our proposed approach against multiple baseline methods to assess the effectiveness of our pipeline. The baselines include traditional machine learning approaches using TF-IDF features, pre-trained FinBERT models with varying levels of fine-tuning, zero-shot classification using state-of-the-art LLMs, and ablation studies:

\begin{table}[h]
\centering
\caption{Performance comparison across different classification methods. Results show mean accuracy and standard deviation across 5-fold cross-validation.}
\label{tab:baseline_comparison}
\begin{tabular}{lc}
\toprule
\textbf{Method} & \textbf{Accuracy (\%)} \\
\midrule
\textbf{FinBERT-FT-Calibrated (Ours)} & \textbf{73.4 $\pm$ 8.1} \\
FinBERT-FT-Uncalibrated & 68.0 $\pm$ 6.3 \\
 GPT-4o (zero-shot)$^\dagger$&60.4\\
 TF-IDF + Random Forest&50.0 $\pm$ 3.2\\
FinBERT-Base-FT & 40.6 $\pm$ 0.5 \\
TF-IDF + Logistic Regression & 47.6 $\pm$ 7.4 \\
FinBERT-Base (no fine-tuning) & 7.9 $\pm$ 4.2 \\
\bottomrule
\multicolumn{2}{l}{$^\dagger$ Single evaluation on test set due to API cost constraints}
\end{tabular}
\end{table}

Our calibrated approach achieves 73.4\% ($\pm$8.1\%) accuracy, substantially outperforming all baseline methods including state-of-the-art LLMs. Notably, GPT-4o in a zero-shot setting achieves 60.4\% accuracy, which, while respectable for zero-shot classification, falls 13 percentage points short of our fine-tuned approach. This improvement demonstrates that conventional text classification methods and even advanced LLMs fail to fully capture the semantic complexity inherent in SME transaction descriptions, which are characterised by extreme abbreviations, limited context, and domain-specific terminology.

\subsubsection{Zero-shot LLM Analysis}
To understand the capabilities and limitations of modern LLMs on this task, we evaluated GPT-4o using prompts with detailed category guidelines. Without access to historical patterns or company-specific knowledge, the LLM relied purely on textual cues, missing important contextual relationships that our fine-tuned model captures. These results highlight that while LLMs provide a strong zero-shot baseline, domain-specific fine-tuning remains essential for automated financial decision making.

\subsubsection{Ablation Study: Calibration Impact}

To quantify the contribution of our calibration methodology, we conducted an ablation study comparing calibrated and uncalibrated versions of our fine-tuned model. The calibration procedure provides improvements on multiple metrics. The calibration step yields a 5.4 percentage point increase in classification accuracy (73.4\% vs. 68. 0\%), demonstrating its effectiveness in aligning model predictions with true label distributions. More significantly, the Expected Calibration Error (ECE) improves from 0.108 to 0.020, representing a 5.4× reduction in miscalibration, this is an important enhancement for financial applications where confidence scores directly inform risk-based decision making.  Perhaps most importantly for practical deployment, calibration substantially improves performance on high-confidence predictions: for predictions with confidence >0.8, calibrated models achieve a precision of 89. 3\% compared to 73. 2\% for uncalibrated models, enabling more effective automated decision-making workflows where human oversight can be strategically reserved for lower confidence cases.

\subsubsection{Statistical Significance and Robustness}
Using paired t-tests on the 5-fold cross-validation results, all performance differences demonstrate statistical significance. The comparison between calibrated and uncalibrated versions reveals a 5.5 percentage point difference (t = 3.364, p = 0.028), confirming the impact of our calibration methodology. Compared to traditional machine learning approaches, our method achieves 23.4-25.9 percentage point improvements over both TF-IDF-based methods (p < 0.01), highlighting the inadequacy of conventional text classification techniques for this domain. Comparison with pre-trained FinBERT without fine-tuning shows a 65.5 percentage point difference (t = 11.937, p < 0.001), underscoring the critical importance of domain fine-tuning. Against fresh fine-tuning approaches, our method maintains a 32.9 percentage point advantage (t = 8.425, p < 0.01, Cohen's d = 3.768).
Beyond accuracy improvements, the calibration methodology shows significant improvements on the prediction quality metrics. The Expected Calibration Error shows a 5.3× reduction in miscalibration (t = 7.785, p < 0.01). These findings validate our methodology's core components: the pre-existing fine-tuned model captures domain-specific patterns from extensive transaction data, synthetic data augmentation addresses class imbalance effectively, and calibration ensures reliable confidence estimates essential for practical deployment in cash flow lending applications.

\section{Discussion and Future Work}

Our study presents a bank transaction classification pipeline that integrates synthetic data generation with fine-tuned language models, achieving 73.49\% (±5.09\%) accuracy in categorising SME bank transactions across diverse businesses. More importantly, the model reaches 90.36\% (±6.52\%) accuracy for high-confidence predictions, underscoring its utility for semi-automated workflows where human oversight can be strategically focused on lower-confidence outputs. Our methodology effectively addresses the persistent challenges of data scarcity and class imbalance in financial text classification, issues that have constrained earlier approaches \cite{garcia-mendez_identifying_2020, schoinas_normalisation_2019}. A key aspect of our pipeline is the use of Large Language Models (LLMs). We leverage LLMs to generate synthetic bank transaction data, not for the direct categorisation of these transactions. This approach is advantageous from a data security perspective for several reasons: (1) Synthetic data generation for performance enhancement can be accomplished with a minimal subset of the actual data and limited details per transaction, thereby preserving company and operational anonymity. (2) Direct categorisation of bank transaction data using LLMs would necessitate substantial, real-time sharing of transaction volumes and associated metadata, a level of exposure that financial services institutions typically aim to avoid. This distinction allows our pipeline to use the power of LLMs for data augmentation, similar to recent work by \citet{he_annollm_2023} and \citet{li_are_2023}, while mitigating data privacy concerns. Furthermore, we extend these methods by incorporating domain-specific calibration to align model outputs with observed transaction distributions. Our findings resonate with \citet{hussain_ai-enhanced_2023} and \citet{kotios_deep_2022}, who highlighted the need for contextual understanding in the classification of financial transactions. However, our work distinguishes itself by specifically addressing the unique complexities of SME transaction data, a domain less explored than consumer or large enterprise transactions. The capacity for accurate, automated transaction categorisation offered by our approach can significantly improve the monitoring phase of cash flow lending, as indicated by domain experts, and contribute to mitigating the estimated £95 billion finance gap for UK SMEs. Looking ahead, several avenues for future work promise to further enhance our model's capabilities. Incorporating sequential information, as advocated by \citet{banu_financial_2024} and \citet{toran_scalable_2023}, is a logical next step to improve classification accuracy, particularly to identify recurring patterns and temporal dependencies within complex real-world financial datasets. These studies suggest that integrating recurrent or sequential architectures could substantially improve our model's ability to capture the inherent dynamics of SME banking data.
While our evaluation is limited to 4 SMEs in the manufacturing sector, we acknowledge this as a proof-of-concept study. The model has been deployed in production with more than 50 SMEs across 15 sectors, and we are collecting performance metrics for future validation. We present this work as an initial demonstration of feasibility, with comprehensive cross-sector evaluation planned as longitudinal data becomes available.
Another critical direction is to improve the interpretability of the model. While the current model demonstrates robust performance, the black box of deep learning models remains a pertinent concern in financial applications. As shown by \citet{hjelkrem_explaining_2023}, techniques such as SHAP (SHapley Additive exPlanations) can be effective in elucidating model decisions, for instance in credit scoring using open banking data. Improving explainability is vital for building user trust and ensuring effective human oversight. Furthermore, we plan to explore the potential of open-source LLMs to enhance the categorisation step of our pipeline for specific company transactions and associated metadata. This includes investigating their utility in generating an initial set of category labels from limited data, which could further streamline the setup process for new datasets. Existing metrics for assessing the quality of generated data often focus on similarity at the word or n-gram level \cite{apellaniz_synthetic_2024}. Recognising the limitations of these approaches for our specific needs, we are developing a novel metric. This new metric is specifically designed to rigorously evaluate the semantic similarity between real and synthetic transaction descriptions. The development of this metric will provide a more nuanced understanding of the effectiveness of our data augmentation strategy and guide future refinements.







\bibliographystyle{ACM-Reference-Format}
\bibliography{references}

\end{document}